\documentstyle[aps,psfig,multicol,amsmath,eqsecnum]{revtex}
\begin{document}
%
\title{Density of states and electron concentration of double heterojunctions subjected to an in-plane magnetic field.}
\author{C. D. Simserides}
\address{INFM and Scuola Normale Superiore, Piazza dei Cavalieri 7, I--56126 Pisa, Italia}
\maketitle

\begin{abstract}
We calculate the electronic states of
Al$_x$Ga$_{1-x}$As/GaAs/Al$_x$Ga$_{1-x}$As double heterojunctions
subjected to a magnetic field parallel to the quasi two-dimensional
electron gas. We study the energy dispersion curves, the density of states,
the electron concentration and the distribution of the electrons in the
subbands.

The parallel magnetic field induces severe changes in the density of states,
which are of crucial importance for the explanation of the
magnetoconductivity in these structures. However, to our knowledge,
there is no systematic study of the density of states under these
circumstances. We attempt a contribution in this direction.

For symmetric heterostructures, the depopulation of the higher subbands,
the transition from a single to a bilayer electron system and the domination
of the bulk Landau levels in the centre the wide quantum well,
as the magnetic field is continuously increased, are presented in the
``energy dispersion picture'' as well as in the
``electron concentration picture'' and in the 
``density of states picture''.
\end{abstract}
\begin{multicols}{2}
\narrowtext

\pagebreak
\section{Introduction}

Althouth the behaviour of the Quasi Two-Dimensional Electron Gas (Q2DEG)
in the presence of a perpendicular magnetic field has been studied
extensively, much less attention has been devoted to the situation
where the magnetic field is applied parallel to the Q2DEG. In the
former case, interesting phenomena e.g. the Shubnikov-de Haas effect
\cite{one} and the integer \cite{two} and the fractional \cite{three}
quantum Hall effects have been observed. In the latter case, electrons
move under the competing influence of the Lorenz force and the force due
to the quantum well confining potential. 

In the presence of an in-plane magnetic field, $B$, single heterojunctions
\cite{four,five,six}, single \cite{seven,eight},
double \cite{eight,nine,ten,eleven,twelve} and
triple \cite{thirteen} square quantum wells,
almost square quantum wells \cite{fourteen,fifteen},
asymmetric square quantum wells \cite{sixteen},
symmetrical wide single quantum wells\cite{seventeen,eighteen}
and superlattices \cite{nineteen} have been considered.

The experimental studies include single heterojunctions \cite{six},
double square quantum wells \cite{ten,eleven,twelve},
triple square quantum wells \cite{thirteen},
wide single quantum wells \cite{eighteen}
and superlattices \cite{nineteen}. The most important experimental
finding \cite{ten,twelve,thirteen,eighteen} is, according to our opinion,
the strong conductance ``oscillations'' due to the variation of the density
of states (DOS), as $B$ is increased. Conductance maxima are identified with
depopulations of local energy dispersion minima, while conductance minima
are identified with van Hove singularities at the chemical potential.
This situation has been encountered in
symmetrical square double \cite{ten,twelve} and
triple \cite{thirteen} quantum wells and in 
symmetrical wide single quantum wells \cite{eighteen}.
While in the cases of square double and triple quantum wells
a simple Tight Binding calculation gave the position of the maxima
and the minima, a self-consistent calculation was indispensable
in the case of symmetrical wide single quantum wells. 

Theoretical studies of the electronic states are usually restricted to
simple analytically solvable potential wells, to Tight Binding Approximation,
or to perturbative approximations. Self-consistent studies are up to now
a few, regarding single heterojunctions \cite{four,five}, 
thin single quantum wells \cite{fourteen} and
symmetrical wide single quantum wells \cite{seventeen}.

In the present work, we use self-consistent calculations to study
Al$_x$Ga$_{1-x}$As/GaAs/Al$_x$Ga$_{1-x}$As wide double heterojunctions
(i.e. a system of two heterojunctions with relatively large distance
between the two interfaces) subjected to an in-plane magnetic field.
Below, we summarize the particular aims of this work. 

Our first aim is to study the density of states when the Q2DEG is
subjected to an in-plane magnetic field. In this case, the DOS is not
a step-like function, as it is with $B$=0. We show that its form undergoes
important changes as $B$ is increased, especially in
wide double heterojunctions where usually many subbands are
present \cite{eikosi,eikosiena}. The self-consistent study of the
electronic states and specifically of the DOS is of great importance
for the explanation of the experimental magnetoconductivity in these
structures. However, up to now, there is no systematic study of the
DOS under these circumstances. We attempt to give a contribution
in this direction.

Our second aim is to study a bilayer electron system, different from the
commonly used symmetrical double square well. Another potentially bilayer
electron system is the symmetrical double heterojunctions, when the well
width is increased a lot, due to the transition from a ``perfect'' square
quantum well to a system of two separated heterojunctions \cite{eikosi}.
In the former structure a high barrier separates the two electron layers.
In the latter structrure the barrier is formed from the redistribution of
the carriers in the well and it is relatively weak. Moreover,
Smr\u{c}ka and Jungwirth \cite{seventeen} have shown, by calculating
the energy dispersion curves in a two-subband situation,
that symmetrical wide single quantum wells can be potentially
bilayer electron systems when the parallel magnetic field is increased.
Here, we present the depopulations of the higher subbands
and the transition from a single layer to a bilayer electron system not
only in the ``energy dispersion picture'', but also in the
``electron concentration picture''. Thus, we calculate
the electron concentration, $n(z)$ and the distribution of the electrons
in the subbands, $n_i(z)$. We also give the ``the density of states''
picture, which is important for the interpretation of
the transport experiments. Moreover, we show in these three pictures
that in the centre of our wide quantum well, as the magnetic field is
further increased, the bulk Landau levels dominate. Finally,
we give an example of an asymmetric heterostructure.

The basic theory is presented in section 2 together with some analogies
to the classical picture. In section 3 we present the theoretical results
for Al$_x$Ga$_{1-x}$As/GaAs/Al$_x$Ga$_{1-x}$As double heterojunctions
and we comment on some interesting features observed.
Our conclusions are summarized in section 4.

\section{Basic Theory}

When a magnetic field, $B$, parallel to the y-axis, is applied to
a three dimensional electron gas, the motion in the xz-plane is
quantized to Landau levels with energy eigenvalues
$E_{xz}= \hbar \omega (i+ \frac {1}{2})$, where $i$ is a
discrete quantum number, $\hbar$ is the reduced Planck constant and
$\omega=\frac{eB}{m^*}$, is the cyclotron frequency.
$m^*$ is the effective mass and q=-e is the electron charge.
If we additionally apply an electric field, $\vec E$, in the z-axis,
$E_{xz}$ depend not only on $i$, but also on the wavevector in the x-axis,
$k_x$.
Specifically, $E_{xz}= \hbar \omega (i+ \frac {1}{2})-\frac {m^*}{2}
(\frac {E}{B})^2-\hbar k_x(\frac {E}{B})$.
In this work we are interested in the configuration with a quantum well
in the z-axis (with or without the electric field in the z-axis)
and the magnetic field in the y-axis. Again, as we discuss below,
$E_{xz}$ depend on both $i$ and $k_x$. However, generally in this case
$E_{xz} = E_i(k_x)$ cannot be expressed analytically and have to be
determined self-consistently. Of course, without a magnetic field,
$E_{xz} = E_i+\frac {\hbar^2 k_x^2}{2m^*}$, where now $i$ is the subband
index. In all the situations described above, the energy eigenvalue
in the y-axis is $E_y = \frac {\hbar^2 k_y^2}{2m^*}$ ,where $k_y$ is the
wavevector in the y-axis. The spin part of the eigenenergy is
$E_{spin} = \pm\frac {1}{2}g^*\mu_BB$, where $g^*$ is
the effective Land\`{e} factor and $\mu_B $ is the Bohr magneton.

Summarizing, in the present configuration, there is a magnetic field
in the y-axis, a quantum well in the z-axis and possibly
an electric field in the z-axis (e.g. an external field due to a gate).
With our choice of axes, the Hamiltonian is:

\begin{equation}
\hat H_{tot}=({\vec p}-q{\vec A})^2/(2m^*)+U(z)+g^*\mu_B{\vec \sigma}
\cdot {\vec B},
\end{equation}

\noindent where $\vec p$ is the momentum operator, $\vec A$ is the vector
potential, $m^*=0.067m_e$ is the GaAs effective mass, $m_e$ is the electron mass,
$\vec \sigma$ is the spin and $\vec B$ is the parallel magnetic field.

\begin{equation}
U(z)=U_{band\;offset}(z)+U_C(z)+U_{XC}(z)+U_E(z),
\end{equation}

\noindent where $U_{band \; offset}(z)$ is the potential energy term due
to the conduction band minima discontinuity, $U_C(z)$ is
the Coulombic potential energy, $U_{XC}(z)$ is
the exchange and correlation potential energy \cite{eikosi}
and $U_E(z)$ is the potential energy due to an electric field applied
in the z-axis e.g. due to a gate.
The magnetic field is applied in the y-axis i.e. ${\vec B}=(0,B,0)$.
For the vector potential we choose ${\vec A}=(Bz,0,0)$ \cite{eikosidio}.
The Hamiltonian becomes:

\begin{equation}
\hat H_{tot}=(\hat p_x-qB\hat z)^2/(2m^*)+\hat p_y^2/(2m^*)+
\hat p_z^2/(2m^*)+U(z)+g^*\mu_B{\vec \sigma} \cdot {\vec B}.
\end{equation}

\noindent We split the spatial and the spin part.
$\Psi({\vec r},{\vec \sigma})=\psi(\vec r) \; \alpha ({\vec \sigma})$
and $E_{tot}=E_{xyz} \pm \frac {1}{2}g^*\mu_BB$.
For the spatial part the envelope function equation writes:

\begin{equation}
[(\hat p_x-qB\hat z)^2/(2m^*)+\hat p_y^2/(2m^*)+\hat p_z^2/(2m^*)+
U(z)]\psi=E_{xyz}\psi.
\end{equation}

\noindent $[\hat p_x,\hat H]=[\hat p_y,\hat H]=0$.
Thus, we look for solutions in the form
$\psi=\frac{1}{\sqrt{S}}\zeta(z) e^{ik_xx} e^{ik_yy}$,
where S=L$_x$L$_y$ is the area of the heterostructure in the xy-plane.
The coordinate y splits from the coordinates x and z.
We have $E_y=\frac {\hbar^2 k_y^2}{2m^*}$, while in the xz plane:

\begin{equation}
\frac{d^2\zeta(z)}{dz^2} +
\frac{2m^*}{\hbar^2}[E_{xz}-\frac{m^*}{2}(\frac{eB}{m^*})^2
(z+\frac{\hbar k_x}{eB})^2-U(z)]\zeta(z)=0. 
\end{equation}

\noindent The non-magnetic part of the potential energy is U(z),
while the magnetic part of the potential energy is
$\frac{m^*}{2}(\frac{eB}{m^*})^2 (z+\frac{\hbar k_x}{eB})^2$.
The center of the magnetic potential energy is the point
$z_0=-\frac{\hbar k_x}{eB}=-\frac{\hbar k_x}{m^* \omega}$.
Thus, the electron is free in the y-axis, but the magnetic field
correlates the motion in the x-axis and the z-axis.
The motion in the xz-plane is characterised by a running wave
$e^{ik_xx}$ and the bound state $\zeta_{i,k_x}(z)$ which
depends on both i and $k_x$.

The energy eigenvalues are:

\begin{equation}
E_{tot}=E_{xz}+E_y \pm \frac{1}{2}g^*\mu_BB=E_i(k_x)+\frac
{\hbar^2 k_y^2}{2m^*} \pm \frac{1}{2}g^*\mu_BB,
\end{equation}

\noindent where, generally $E_i(k_x) \ne E_i(-k_x)$.

The density of states is:

\begin{equation}
n({\mathcal E})=\sum_{i,k_x,k_y,\sigma}
\delta ({\mathcal E}-E_{i,k_x,k_y,\sigma})=\sum_{i,k_x} n_{i,k_x}
({\mathcal E}),
\end{equation}

\noindent where:

\begin{equation}
n_{i,k_x} ({\mathcal E})=\sum_{k_y,\sigma}
\delta({\mathcal E}-E_{i,k_x,k_y,\sigma})=
2\sum_{k_y}\delta({\mathcal E}-E_{i,k_x}-\frac{\hbar^2 k_y^2}{2m^*}).
\end{equation}

\noindent We have used the symbolism $E_{i,k_x} \equiv E_i(k_x)$.
Intergrating over $k_y$, Eq. 8 writes:

\begin{equation}
n_{i,k_x} ({\mathcal E})=2\frac{L_y\sqrt{2m^*}}{4\pi\hbar}
 \frac{1} {    \sqrt {{\mathcal E}-E_{i,k_x}}\      } 
\cdot \Theta({\mathcal E}-E_{i,k_x}),
\end{equation}

\noindent where $\Theta$ is the step function.
We must note here that the DOS is not a step-like function,
as it is with zero magnetic field.

The electron concentration is:

\begin{equation}
n(\vec r)=\sum_{i,k_x} n_{i,k_x}(\vec r),
\end{equation}

\noindent where:

\begin{equation}
n_{i,k_x}(\vec r)=\int_{-\infty}^{+\infty}
 d{\mathcal E}n_{i,k_x} ({\mathcal E})
f_0({\mathcal E})|\psi_{i,k_x}(\vec r)|^2.
\end{equation}

\noindent $f_0({\mathcal E})$ is the Fermi-Dirac distribution function and
$\psi_{i,k_x}(\vec r)$ is the three-dimensional envelope function.
Thus, at finite temperature , $T$:

\begin{eqnarray}
\lefteqn{
n_{i,k_x}(\vec r)=2 \frac {\sqrt{2m^*}}{4 \pi \hbar L_x}|\zeta_{i,k_x}(z)|^2}
\nonumber\\
& & {} \int_{0}^{+\infty} d{\alpha} \frac {1}{\sqrt{\alpha}}\;
\frac {1}{1+exp(\frac {\alpha+E_{i,k_x}-\mu (T)}{k_BT})},
\end{eqnarray}

\noindent where $\mu (T)$ is the chemical potential and k$_B$ is the Boltzmann
constant. Using Eq. (12), Eq. (10) becomes:

\begin{eqnarray}
\lefteqn{n(z)=\sum_i n_i(z)=
\sum_i \sqrt{\frac {2m^*}{\hbar^2}} \frac {1} {(2 \pi)^2}
\int_{-\infty}^{+\infty}dk_x|\zeta_{i,k_x}(z)|^2}
\nonumber\\
& & {} \int_{0}^{+\infty} d{\alpha} \frac {1}{\sqrt{\alpha}}\;
\frac {1}{1+exp(\frac {\alpha+E_{i,k_x}-\mu (T)}{k_BT})}.
\end{eqnarray}

\noindent Therefore, the sheet electron concentration, is:

\begin{eqnarray}
\lefteqn{N_s=\sum_i N_i=\sum_i \sqrt{\frac {2m^*}{\hbar^2}} \frac {1} {(2 \pi)^2}
\int_{-\infty}^{+\infty}dk_x}
\nonumber\\
& & {} \int_{0}^{+\infty} d{\alpha} \frac {1}{\sqrt{\alpha}}\;
\frac {1}{1+exp(\frac {\alpha+E_{i,k_x}-\mu (T)}{k_BT})}.
\end{eqnarray}

For a Hamiltonian like that in Eq. (1),
$m^* {\vec v}={\vec p}-q{\vec A}$ \cite{eikosidio}, which in our case
becomes $m^*\hat v_x=\hat p_x+eB \hat z$, $m^*\hat v_y=\hat p_y$
and $m^*\hat v_z=\hat p_z$. Thus, after a little algebra we obtain
for the accelaration and the force operators in the x, y and z axes,
respectively: $\hat a_x = + \omega \hat v_z$,
$\hat F_x = + m^* \omega \hat v_z$, $\hat a_y = 0$,
$\hat F_y = 0$ and $\hat a_z =
- \omega \hat v_x -\frac {1}{m^*} \frac {\partial U(\hat z)}
{\partial \hat z}$, $\hat F_z = - m^* \omega \hat v_x - \frac 
{{\partial U(\hat z)}}{\partial \hat z}$.
So, in the y-axis there is no force on the electrons,
in the x-axis there is only the Lorenz force, while in the z-axis there
is apart from the Lorenz force, the force due to the quantum well
confining potential.

When there is no quantum well ($U(z)=0$) the quantities
$\hat z_0= - \frac {\hat p_x}{eB}$, which corresponds to the z-coordinate
of the center of the classical cyclic orbit, and
$\hat x_0= \frac {\hat p_z} {eB}+\hat x$, which corresponds to the
x-coordinate of the center of the classical cyclic orbit, are constants
of the motion. When $U(z)=0$, the quantity
$\hat r_c^2=\frac {(\hat p_x+eB\hat z)^2+\hat p_z^2}{{m^*}^2 \omega^2}$,
which corresponds to the square of the radius of the classical cyclic orbit
is also a constant of the motion. Thus, when $U(z)$ can be ignored,
electrons describe the well-known spiral motion.

The algorithm used to solve self-consistently the equations above,
is divided in the following steps.
({$\alpha$}\'{}) We input an initial guess for the non-magnetic potential
energy, $U_{in}(z)$.
({$\beta$}\'{}) We solve the envelope function Eq. (5) for each i and for
each $k_x$ to obtain $\zeta _{i,k_x}(z)$ and $E_{i,k_x}$. Care should be
taken in this step, to include all possible i and all possible $k_x$ which
contribute to the electron concentration. Thus, we start with many subbands
and with a wide range of k$_x$. This means that Eq. (5) must be solved
{\it many} times.
({$\gamma$}\'{}) $\mu (T)$ can be calculated from charge
neutrality \cite{eikosi,eikosiena}, using Eq. (14).
({$\delta$}\'{}) Thus, we can calculate, from Eq. (13)
$n(z)$ and $n_i(z)$ and therefore $U_{XC}(z)$ \cite{eikosi}.
({$\epsilon$}\'{}) Now the charge density is known and it is used to solve
the Poisson equation numerically \cite{eikosi}, to obtain $U_C(z)$.
We suppose that $\frac {dU_C}{dz}({\it bulk})=0$, because there is no
net charge in the bulk material. We take into account the different
dielectric constants of GaAs and Al$_x$Ga$_{1-x}$As \cite{eikosi}.
Finally, we choose $U_C({\it left \; bulk})=-U_0$, where $U_0$ is
the value of the conduction band minima discontinuity.
$U_{XC}({\it bulk})=0$, because the envelope functions decay into the
Al$_x$Ga$_{1-x}$As barriers. Thus, $U({\it left \; bulk})=0$.
All the structures are long enough in the z-axis, so that bulk
conditions prevail before the end of the Al$_x$Ga$_{1-x}$As barriers.   
({$\varsigma$}\'{}) The output non-magnetic potential energy,
$U_{out}(z)$, can now be calculated from Eq. (2).
({$\zeta$}\'{}) If $U_{out}(z)$ is ``very close'' to $U_{in}(z)$,
we have finished. Otherwise, we mix $U_{out}(z)$ and $U_{in}(z)$
to construct the new $U_{in}(z)$ and we return to step
({$\beta$}\'{}) \cite{eikosi}.

Finally, we notice that since the envelope functions depend on both i
and $k_x$, a quantitative calculation of the conductivity will involve
tedious algebra, because the scattering matrix elements will depend on
$k_x$, too. This complication emerges also in the calculation
of the screening. For this reason, although some attemps have already
been made \cite{eikositria}, a well established transport theory for
a Q2DEG with an in-plane magnetic field has not been developed yet.

\section{Results and Discussion}

We apply our treatment to the case of a symmetrical $\delta$-doped
Al$_x$Ga$_{1-x}$As/GaAs/Al$_x$Ga$_{1-x}$As double heterojunction,
in the presence of a parallel magnetic field, from 0T up to 20T.
We choose a symmetrical structure because in this case we can observe
most clearly the variation of the electronic states induced by
the magnetic field in the ``energy dispersion picture'',
in the ``electron concentration picture'' and in the
``density of states picture''. 
The structure consists of a 280{\AA} undoped Al$_{0.25}$Ga$_{0.75}$As layer,
a Si $\delta$-doped Al$_{0.25}$Ga$_{0.75}$As layer
(0.28 $\times$ $10^{12}$ cm$^{-2}$),
an undoped 250{\AA} Al$_{0.25}$Ga$_{0.75}$As spacer,
a 600{\AA} undoped GaAs well, an  undoped 250{\AA} Al$_{0.25}$Ga$_{0.75}$As
spacer, a Si $\delta$-doped Al$_{0.25}$Ga$_{0.75}$As layer
(0.28 $\times$ $10^{12}$ cm$^{-2}$) and a 280{\AA}
undoped Al$_{0.25}$Ga$_{0.75}$As layer. All layers are assumed to have
a slight unintentional acceptor doping of 5 $\times$ $10^{14}$ cm$^{-3}$.
We suppose that the sample has been illuminated and therefore all the
donors are ionized. This is done because we want to study the effect of
the magnetic field under the condition of constant sheet electron
concentration. Although our treatment is applicable to any temperature,
we will apply it to $T$ = 4.2K. This is done because the experiments
are usually performed at or below $T$ = 4.2K.
These material and structural parameters result in a sheet
electron concentration, N$_s$= 0.54 $\times$ $10^{12}$ cm$^{-2}$. 

First we will describe the evolution of the changes induced by the
magnetic field to the energy dispersion curves, $E_{i}(k_x)$.
The situation is described in the lower parts of Fig. 1.
In this particular structure, for $B=0$, due to the large well width,
the ground state subband and the first and second excited subbands are
populated, with sheet electron concentrations $N_0$= 0.238
$\times$ $10^{12}$ cm$^{-2}$, $N_1$= 0.233 $\times$ $10^{12}$ cm$^{-2}$
and $N_2$=0.069  $\times$ $10^{12}$ cm$^{-2}$, respectively.
Initially, as the magnetic field is increased, depopulation
of the higher subbands is predicted. The second excited subband is
depopulated at $B \simeq$ 5T and the first excited subband at $B \simeq$ 7T.
Increasing the magnetic field up to 7T, the shapes of the $E_{i}(k_x)$
dispersion curves also change. While the upper subbands remain
almost parabolic, the first excited subband, and most obviously
the ground state subband undergo important changes, developing gradually
local maxima at $k_x=0$ instead of local minima at $B$ = 0T.
As can be seen from the lower parts of Fig. 1 this also happens to the
other excited subbands for larger values of the magnetic field.
For $B>$ 7T, only the ground state subband is populated.
At this point the $E_{0}(k_x)$ dispersion curve is continuously below
the chemical potential in the range
$k_x=[-4 \times 10^8,+4 \times 10^8]$ m$^{-1}$.
This means that the system is still a single layer one.

At higher magnetic fields, a transition from a single to a bilayer electron
system occurs. This transition has approximately been achieved at
$B$ = 12T as can be seen from the lower part of Fig. 1c, but the complete
separation of the two layers is achieved at $B$ = 20T
(see Fig. 2 where the electron concentrations are presented).
During this procedure, the energy separation of the unoccupied states
(those with small $|k_x|$) becomes $\hbar \omega$.
This is due to the fact that the well width is very large and therefore
in the central region of the well, as the magnetic confinement
overcomes the well confinement, the bulk Landau levels dominate.
This has also been predicted for square, analytically solvable,
quantum wells when the well width is large enough \cite{sixteen}.

In the case of a square quantum well, the behaviour of the $E_{i}(k_x)$
curves is determined by the competition of the {\it well width} and the
{\it magnetic length}, $l_B=\sqrt{\hbar/(eB)}$ \cite{sixteen}.
When the well width is smaller than the magnetic length, spatial
quantization dominates. The energy levels can be roughly classified
into two types, namely, {\it confined} states and {\it extended} states.
In this {\it specific case} the confined states in the quantum well
increase parabolically as a function of $k_x$ \cite{seven,fourteen},
while the extended states have an oscillating form with an ``average''
separation of $\hbar \omega$ \cite{seven}. However, as the well width
or the magnetic field is increased, this behavior changes.
Finally, when the well width is larger than the magnetic length,
the electron orbits are governed by the Lorentz force and electrons
basically describe spiral motion. At this point the energy dispersion
curves are flat with a separation of $\hbar \omega$.

In reference \cite{fourteen}, the author, studing thin single quantum
wells and taking as the growth axis the z-axis and the magnetic field
in the x-axis, bypasses the dependence of the electronic states on the
in-plane wavevector in the y-axis (perpendicular to B), using only $k_y=0$.
This is done in order to override the large numerical cost of the general
case. It is evident from the lower parts of Fig.1 that such an
approximation cannot be applied in our case because of the strong
depedence of the electronic states on this wavevector. Moreover,
for high enough values of the magnetic field the states with this
wavevector are not occupied.

The ``density of states picture'' is given in the upper parts of Fig. 1.
We observe that although for $B$ = 1T (Fig. 1a) the DOS is almost step-like,
there is already a peak due to the fact that $E_0(k_x)$ has already
developed a local maximum at $k_x$=0, instead of the local minimum at
$B$=0T. This corresponds to a van Hove singularity,
since $\frac {dE_0(k_x)}{dk_x}>$ 0 as we approach the critical point
from below and $\frac {dE_0(k_x)}{dk_x}<$ 0 as we approach
the critical point from above. The DOS of the first excited subband
is not a ``perfect step'', because $E_1(k_x)$ is not exactly parabolic.
The DOS for the second and the third excited subbands are
``perfect steps'' because $E_2(k_x)$ and $E_3(k_x)$ are parabolic.
We can also see that we have three populated subbands.

In the upper part of Fig. 1b we present the DOS for $B$ = 7T.
Clearly, we can observe the depopulation of the first excited subband.
Therefore, at this point, as we increase the magnetic field,
the conductivity of the structure increases abruptly, due to the abrupt
decrease of the DOS at the chemical potential. We also observe that the
total DOS is not step-like and that the second and the third excited
subbands are not exactly parabolic. Moreover, since $E_0(k_x)$ and
$E_1(k_x)$ have developed local maxima at $k_x$=0, there are two peaks
in the DOS, corresponding to the two van Hove singularities.

In the upper part of Fig. 1c we present the DOS for $B$ =12T.
There is a van Hove singularity at the chemical potential, due to the
local maximum of $E_0(k_x)$ at $k_x$=0. Therefore, at this point,
as we increase the magnetic field, the conductivity of the structure
decreases abruptly due to the abrupt increase of the DOS at the chemical
potential. The total DOS indicates that at the center of the well the bulk
Landau levels start to develop. All $E_i(k_x)$ have already developed
local maxima at $k_x$=0 and the energy separation of successive subbands
for small $|k_x|$ is close to $\hbar \omega$.

In the upper part of Fig. 1d we present the DOS for $B$ = 20T.
The form of the total DOS stems from the combination of two factors,
i.e. as we move in the energy axis to higher energies:
(a) From the two local minima of $E_i(k_x)$ up to the local maximum
of $E_i(k_x)$ the bilayer electron system dominates.
(b) From the local maximum of $E_i(k_x)$ up to the local minima
of $E_{i+1}(k_x)$ the bulk Landau levels dominate.
In this region the DOS has the form
$constant \times \sum_i (\mathcal E -
\hbar \omega (i + \frac {1} {2}))^{- \frac {1} {2}}$ =
$constant' \times \sum_{i,k_y} \delta (\mathcal E - 
\hbar \omega (i + \frac {1} {2})- \frac {\hbar^2 k_y^2}{2m^*})$,
which is the DOS of a free particle in the y-axis together with
an harmonic oscilator in the xz-plane. The energy separation of
successive subbands for small $|k_x|$ is equal to $\hbar \omega$. 

Fig. 2 presents the variation of the electron concentration,
$n(z)$ and of the population of the subbands, $n_i(z)$, as we increase
the magnetic field from 0T to 20T. The depopulation of the second
excited subband at $B \simeq$ 5T and of the first excited subband
at $B \simeq$ 7T can also be seen in this
``electron concentration picture''.
Inspection of Fig. 2 reveals that apart from these depopulations,
the form of $n(z)$ changes even from 1T to 7T,
with a little bigger separation of the two parts of $n(z)$.
This separation increases with the increase of the magnetic field.
At 12T there are still electrons in the middle of the well.
The division into two parts is complete at 20T.
This means that the ``electron concentration picture''
gives a more precise depiction of the transision to a bilayer system
than the ``energy dispersion picture''. 

We finally give an example of an asymmetric heterostructure.
The structure consists of a 700{\AA} undoped
Al$_{0.25}$Ga$_{0.75}$As layer, a 50{\AA}
Si-doped Al$_{0.25}$Ga$_{0.75}$As layer
(2 $\times$ $10^{18}$ cm$^{-3}$), an undoped
50{\AA} Al$_{0.25}$Ga$_{0.75}$As spacer, a 600{\AA} undoped GaAs well,
an undoped 200{\AA} Al$_{0.25}$Ga$_{0.75}$As spacer,
a 50{\AA} Si-doped Al$_{0.25}$Ga$_{0.75}$As layer
(1 $\times$ $10^{18}$ cm$^{-3}$) and a 600{\AA} undoped
Al$_{0.25}$Ga$_{0.75}$As layer. We suppose, again, that all
the layers have a slight unintentional acceptor doping of
4 $\times$ $10^{14}$ cm$^{-3}$ and that the sample has been illuminated
so that all the donors are ionized. $T$ = 4.2K. These material
and structural parameters result in a sheet electron concentration,
N$_s$= 1.491 $\times$ $10^{12}$ cm$^{-2}$. For $B$ = 0T,
there are four populated subbands with sheet electron concentrations
$N_0$= 0.742 $\times$ $10^{12}$ cm$^{-2}$,
$N_1$= 0.406 $\times$ $10^{12}$ cm$^{-2}$, $N_2$=0.229 
$\times$ $10^{12}$ cm$^{-2}$ and $N_3$=0.114 $\times$ $10^{12}$ cm$^{-2}$,
respectively. 

In Fig. 3 we present the energy dispersion curves, $E_{i}(k_x)$
(lower part) and the density of states (upper part), for $B$ = 5T.
We notice that for this asymmetric heterostructure
$E_i(k_x) \ne E_i(-k_x)$. The populations of the subbands are
now $N_0$= 1.042 $\times$ $10^{12}$ cm$^{-2}$,
$N_1$= 0.324 $\times$ $10^{12}$ cm$^{-2}$, $N_2$=0.110 
$\times$ $10^{12}$ cm$^{-2}$ and $N_3$=0.015
$\times$ $10^{12}$ cm$^{-2}$, respectively. 
We can observe the ``transposition and the anticrossings of the
parabolas'' which result in a complicated form for the DOS.
We can also see that there are two different van Hove singularities
which give the peaks in the DOS. 

Generally, both in the symmetrical and in the asymmetrical case,
the van Hove singularities are not simply saddle points because
the $E_i(k_x)$, as we approach the critical points, are not of
the form $-\alpha k_x^{2}$, $\alpha >$0. The exact form of the
dispersion curves is obtained from the self-consistent calculation.
Anyway, as we increase the magnetic field, whenever the chemical
potential is identified with a van Hove singularity, the conductivity
of the structure will decrease abruptly. On the contrary, whenever
there is a depopulation of a local energy dispersion minimum,
due to the decrease of the DOS at the chemical potential,
the conductivity will increase abruptly.

Similar results are obtained for other values of the well width.
It is the competition between the magnitude of the magnetic
field and the spatial quantization - together with the influence
of the number of electrons - that determines the overall behaviour
of the system. Extensive comparison with experiment will be presented
in a forthcoming paper.

\section{Summary}

We have calculated self-consistently the energy dispersion curves,
the density of states, the electron concentration and the distribution
of the electrons in the subbands,
for Al$_x$Ga$_{1-x}$As/GaAs/ Al$_x$Ga$_{1-x}$As double heterojunctions
subjected to an in-plane magnetic field.

We have systematically studied the important changes in the density
of states, induced by the variation of the in-plane magnetic field.
We have pointed out that these changes are of crucial importance for
the explanation of the magnetoconductivity experiments.

In the case of symmetric heterostructures, we have demonstrated in the
``energy dispersion picture'', in the ``electron concentration picture''
and in the ``density of states picture'' the depopulation of the higher
subbands, the transition from a single to a bilayer electron system and
the domination of the bulk Landau levels in the centre
the wide quantum well, as the magnetic field is continuously increased.
We have also given an example of an asymmetric heterostructure.

The author wishes to thank Prof. Fabio Beltram and Dr. Vincenzo Piazza
for many useful discussions and for the motivation of this work.

\end{multicols}
\pagebreak


\begin{figure}
\vspace*{0.0cm}
\hspace*{0.4cm}\psfig{file=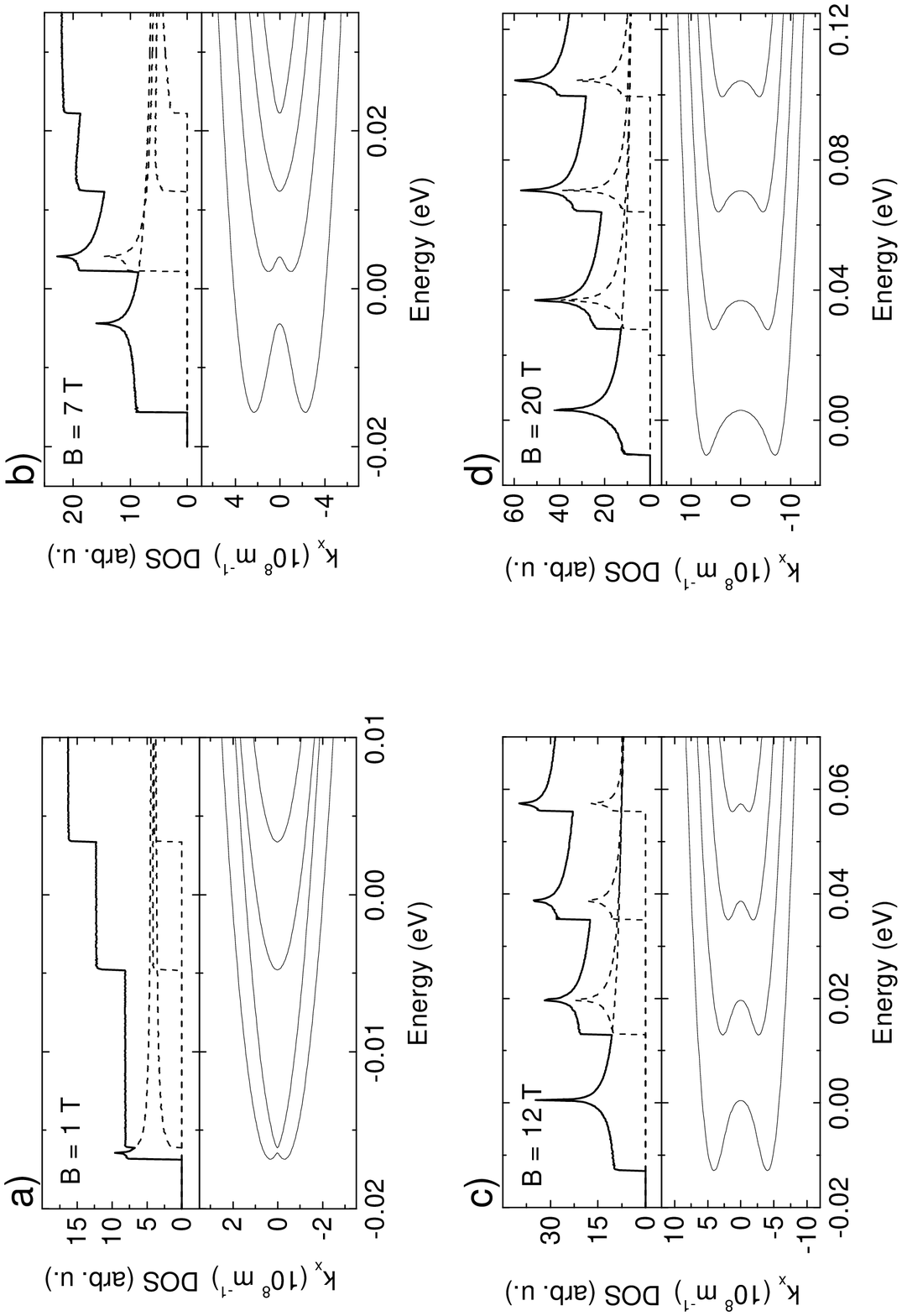,height=6cm}
\vspace*{0.0cm}
\pagebreak
\caption
{{\it Symmetric heterostructure}. The energy dispersion curves,
$E_i(k_x)$, i=0,1,2,3 (lower parts) and the density of states (upper parts)
drawn with a common horizontal energy axis for (a)$B$=1T, (b)$B$=7T,
(c)$B$=12T, and (d)$B$=20T. The DOS is in arbitrary units.
The chemical potential is identified with the zero energy.
The dashed curves represent the DOS of each subband, $n_i({\mathcal E})$,
while the bold continuous curve represents the total DOS, $n({\mathcal E})$.}
\end{figure}
\pagebreak

\begin{figure}
\psfig{file=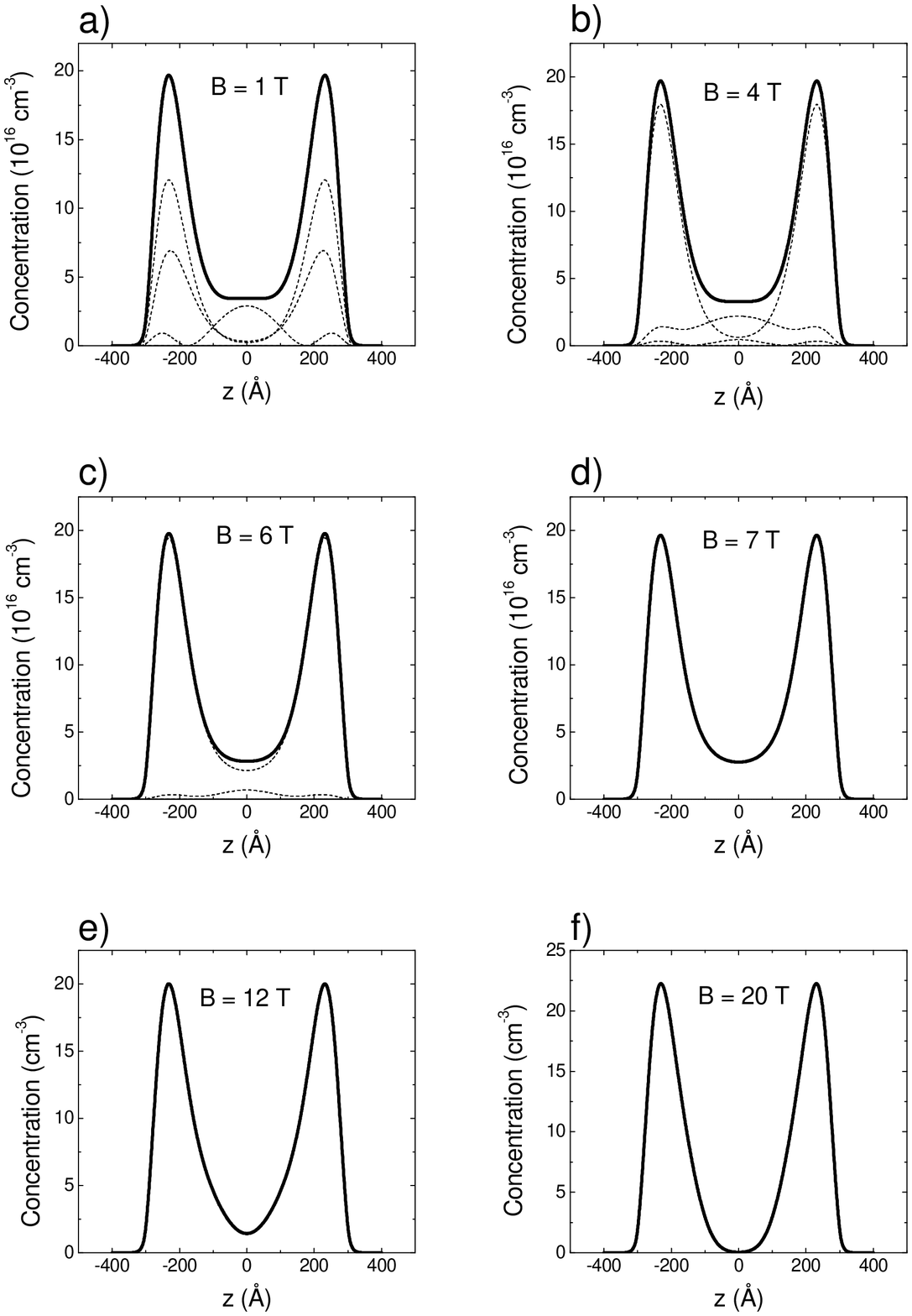,height=16cm}
\pagebreak
\caption
{{\it Symmetric heterostructure}.
The electron concentration, $n(z)$ (bold continuous curve)
and the population of the subbands, $n_i(z)$ (dotted curves)
for (a)$B$=1T, (b)$B$=4T, (c)$B$=6T, (d)$B$=7T, (e)$B$=12T, and (f)$B$=20T.}
\end{figure}
\pagebreak

\begin{figure}
\psfig{file=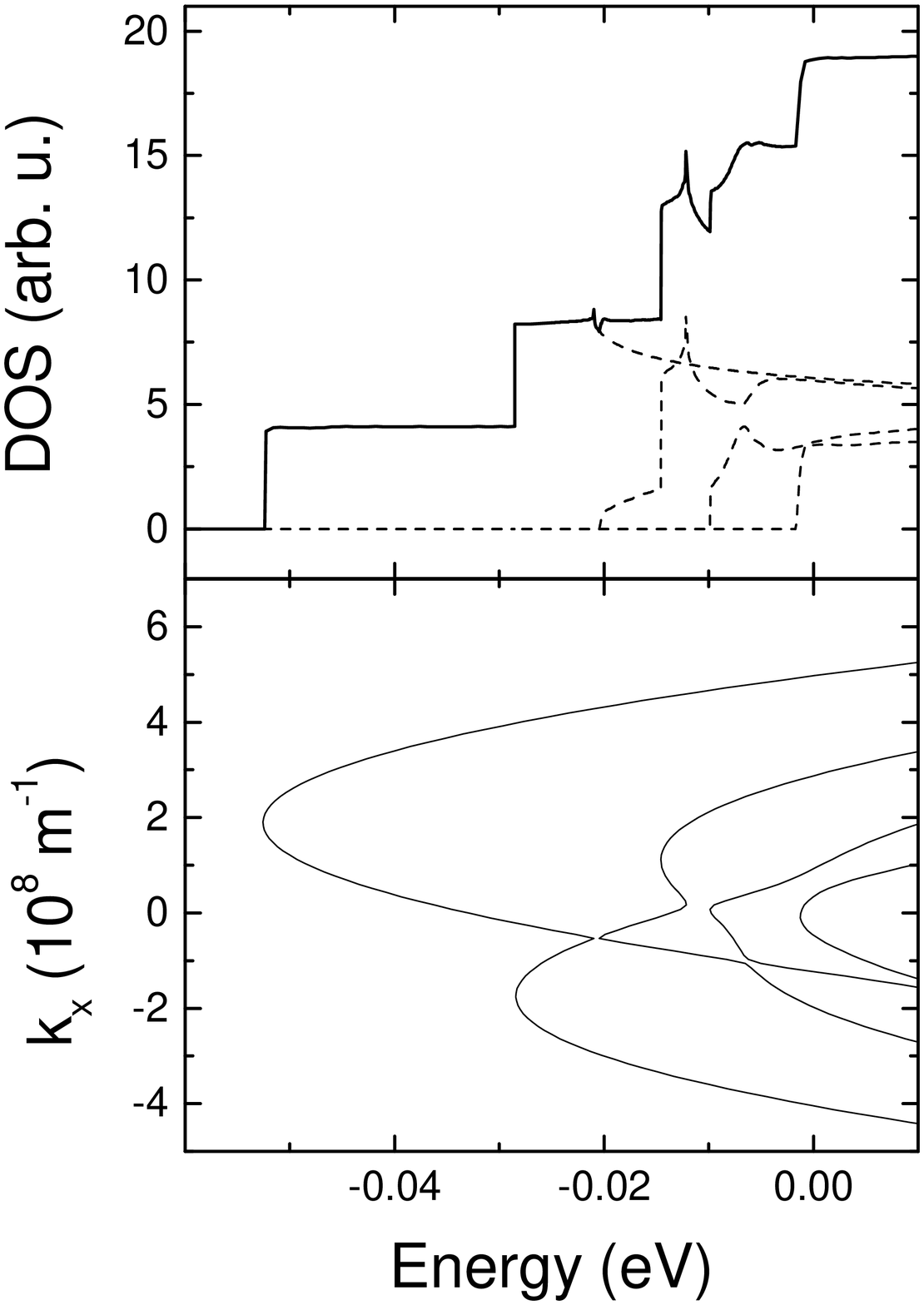,height=16cm}
\pagebreak
\caption
{{\it Asymmetric heterostructure}. The energy dispersion
curves, $E_i(k_x)$, i=0,1,2,3 (lower part) and the density of states
(upper part) drawn with a common horizontal energy axis for $B$=5T.
The DOS is in arbitrary units. The chemical potential is identified
with the zero energy. The dashed curves represent the DOS of each subband
$n_i({\mathcal E})$, while the bold continuous curve represents
the total DOS, $n({\mathcal E})$.}
\end{figure}

\end{document}